\documentclass[conference,compsoc]{IEEEtran}

\usepackage{booktabs}
\usepackage{graphicx}
\usepackage{tabularx}
\usepackage{amsmath}
\usepackage{amssymb}
\usepackage{url}
\usepackage{tikz}
\usetikzlibrary{arrows.meta}
\usepackage[dvipsnames]{xcolor}

\begin{document}

\title{Early Detection of Distributed Backdoors in\\Multi-Agent LLM Systems: A Characterization Study}

\author{\IEEEauthorblockN{Diego Fernandez Arias, Dev Prashant Mistry, Ren Wang, and Yibo Hu}
\IEEEauthorblockA{Illinois Institute of Technology\\
Chicago, Illinois, USA\\
\texttt{\{dfernandezarias,dmistry3\}@hawk.illinoistech.edu}\\
\texttt{\{rwang74,yhu89\}@illinoistech.edu}}}

\maketitle
 
\begin{abstract}
Multi-agent LLM systems can be attacked by a payload that no single agent ever holds in full: several poisoned tools each hide one encrypted fragment, spreading them across several agents, and an external step reassembles and executes them after the run. Per-step safety checks that judge each action in isolation may fail to recognize the complete distributed payload. We investigate how early such an attack can be detected while the run is still unfolding, and how robustly it can be caught once its most obvious cues are stripped away. We build a working instance on a hierarchical multi-agent system, run it under benign and attacked conditions across five language models and two tool environments, and record when each fragment is injected and when the payload is assembled and executed. Almost no run is flagged before its first fragment is injected; once injection begins, a prefix detector flags $99.5\%$ of successful attacks with a median of twelve steps remaining and an out-of-fold false-alarm rate of about $1\%$ on uninjected runs, higher on runs that carried fragments but never assembled. Because assembly occurs only after the run, these alarms could enable an abort before assembly on nearly every successful attack. We also test a detector that sees only the current observation, to ask whether earlier steps are needed. When the current observation carries an obvious clue, one observation is often enough; when that clue is removed, using earlier steps can help. Generic zero-shot and behavior-trained detectors fail to separate unsafe from safe runs; the detectors that do work lean in part on removable surface cues, chiefly the ciphertext's length and entropy. Take those cues away and the detector fires later, and carrying it from one tool environment to the other becomes much harder. A model fine-tuned on the raw text recovers part of the loss. \footnote{Code and a sanitized version of the trajectory corpus will be released at \url{https://github.com/yibo-hu-lab/distributed-backdoor-early-warning}.}

\end{abstract}

\begin{figure*}[!t]
\centering
\includegraphics[width=\textwidth, keepaspectratio]{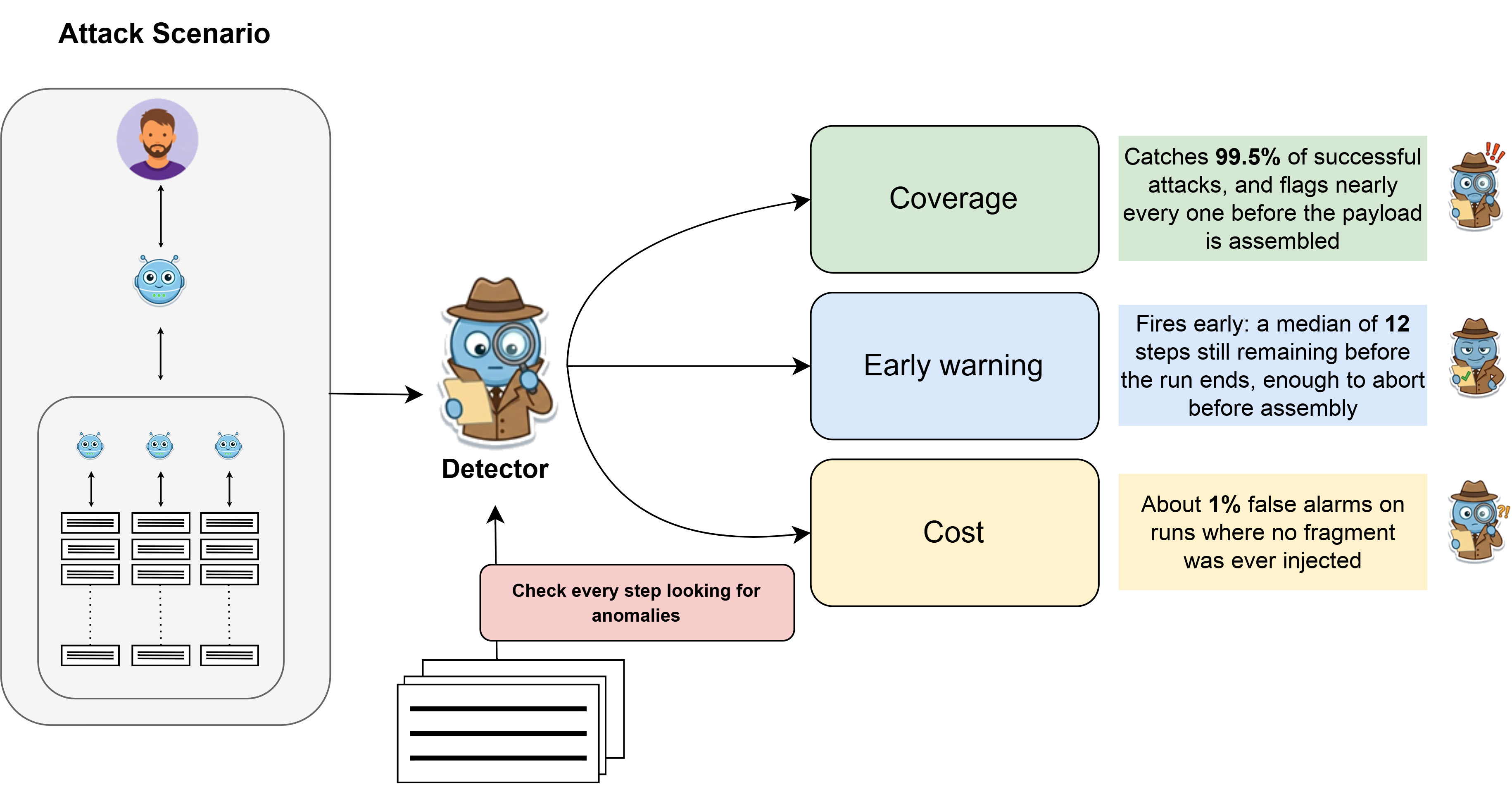}
\caption{The early-warning result at a glance. A prefix detector reading only the run so far catches $99.5\%$ of successful attacks at a median of twelve steps before the run ends, leaving time for an abort before the external assembler reconstructs the payload, at an out-of-fold false-alarm rate of about $1\%$ on uninjected runs. }
\label{fig:teaser}
\end{figure*}

\section{Introduction}
\label{sec:intro}

Large language models are increasingly deployed inside agentic frameworks that plan, call external tools, retrieve documents, and coordinate with one another to carry out multi-step tasks~\cite{yao2023react,wu2023autogen}. Each of these interactions is also an entry point for an attacker. Recent work has shown that an agent can be compromised through almost any component it depends on: its prompt~\cite{xiang2024badchain}, its fine-tuning data~\cite{wang2024badagent}, its retrieval store~\cite{zou2025poisonedrag,cheng2024trojanrag}, or the store of past experiences it draws on~\cite{chen2024agentpoison}. 

Attacks of this kind are \emph{backdoors}. A backdoor plants a hidden trigger so the system behaves normally on ordinary inputs but produces attacker-chosen behavior when a secret condition is met~\cite{zhao2024backdoorsurvey}; in an agent, the trigger need not live in the model weights, and can be delivered through any of the channels above~\cite{deng2024aiagents}. A body of defenses has grown up to catch them: filtering anomalous or high-perplexity tokens before they reach the model, reconstructing candidate triggers to test an input, and inspecting a model's internal activations for the signature of a poisoned response~\cite{zhao2024backdoorsurvey}. What these defenses have in common is that each judges one artifact in isolation.

That is the wrong granularity for an attack that is spread across a run. Collaborative Shadows~\cite{zhu2025collabshadows} makes the point concrete: it hides an encrypted payload by splitting it into fragments, placing one fragment in each of several tool observations across several agents, and reassembling and executing it only after the system has finished. No single fragment reveals the complete malicious payload, so a per-step check sees no complete attack. This makes monitoring the full run a natural defense. The payload is reconstructed from many steps, but an alarm may not need them all, so we measure both when the alarm arrives and how much the earlier steps add over the current observation alone.

We ask a single question: watching such a run unfold step by step, how early can we tell that it is going bad? To study it we built a working instance of the attack on a hierarchical multi-agent system and ran it under benign and attacked conditions across five language models and two tool environments, producing the evaluation corpus for our experiments. We score a detector on growing prefixes of the logged run and ask how early a run can be flagged once it turns malicious. We address three main questions:
\begin{itemize}
 \item After the first malicious fragment is injected, how early can we flag the run as unsafe? Is there enough time to abort before assembly and execution?
 \item What does that early warning cost in false alarms on safe runs, and how does the warning lead-time trade off against the confidence we demand of the detector?
 \item How much of that early warning rests on removable surface cues of the attack rather than on its distributed structure, and how much of it is already available from a single observation? And how reliably does the remaining signal hold once we strip those cues away and move across tool environments?
\end{itemize}

Our central finding is that a distributed attack of this kind can be caught early, but only once it leaves a trace. Before the first payload fragment is injected there is almost nothing for a detector to react to; the meaningful question is how quickly it responds once a fragment appears. Once injection begins, a simple classifier reading only the run so far flags $99.5\%$ of successful attacks with a median of twelve steps remaining before the run ends, at an out-of-fold false-alarm rate of about $1\%$ on uninjected runs. Because assembly occurs only after the agents finish, this creates an actionable window in which almost every successful attack can still be aborted. Detection is therefore a race against assembly, and in this setting that race is winnable. Earlier steps matter less than we expected. A detector that sees only the current observation catches as many attacks as one that reads the whole run. The prefix detector pulls ahead only after we hide what makes the injected text stand out.

How detectable the attack is, however, depends on where the backdoor lives and what signal it leaves behind. An encrypted payload delivered through tool observations leaves a statistical trace in the run, whereas a model backdoored during fine-tuning may leave none. Our detectors are accordingly tailored to the tool-injection family we study, and generic zero-shot and behavior-trained trajectory monitors do not transfer to it. The warning relies heavily on removable payload cues, chiefly ciphertext length and entropy. Take those cues away and the detector fires later, and carrying it from one tool environment to the other becomes much harder. A model fine-tuned on the raw text recovers part of the loss.

The external post-run assembler provides explicit landmarks for fragment injection, assembly, and execution, allowing us to measure not only whether an attack is detected, but whether the warning would leave time to abort before reconstruction. We evaluate this warning window under unseen tasks, domain shift, and reduced surface cues.

This work makes three main contributions: 
\begin{itemize}
 \item \textbf{An instrumented testbed for distributed backdoors.}
 We build a working hierarchical multi-agent system under a distributed payload attack and collect trajectories across five language models and two tool environments, with explicit landmarks for fragment injection, detection, assembly, and execution.

 \item \textbf{An actionable early-warning result.}
 We show that an online prefix detector flags almost every successful attack with a median of twelve steps still to run, and we report its false alarms on uninjected runs separately from its false alarms on runs that carried fragments but failed to assemble.

 \item \textbf{A characterization of what makes that warning possible and why it is brittle.}
 The tested behavior-oriented baselines do not separate unsafe from safe runs. The successful feature detector instead relies on removable payload cues such as length and entropy. Take those cues away and it fires later, and carrying it from one tool environment to the other becomes much harder. We also show when earlier steps help: one observation is usually enough while the injected text still stands out, and the prefix detector pulls ahead only once length is withheld.
\end{itemize}

\section{Related Work}
\label{sec:related}

\begin{table*}[t]
\centering
\small
\setlength{\tabcolsep}{4.5pt}
\renewcommand{\arraystretch}{1.15}
\caption{Comparison with the closest prior work.}
\label{tab:related_work_comparison}
\resizebox{\textwidth}{!}{%
\begin{tabular}{lcccccc}
\toprule
\textbf{Work}
& \textbf{Distributed}
& \textbf{Executed}
& \textbf{Online prefix}
& \textbf{Attack}
& \textbf{Pre-assembly}
& \textbf{Surface-cue} \\
& \textbf{across agents}
& \textbf{payload}
& \textbf{monitoring}
& \textbf{landmarks}
& \textbf{timing}
& \textbf{robustness} \\
\midrule
Collaborative Shadows~\cite{zhu2025collabshadows}
& \checkmark & \checkmark & -- & -- & -- & -- \\

SkillTrojan~\cite{feng2026skilltrojan}
& -- & \checkmark & -- & -- & -- & -- \\

AgentForesight~\cite{zhang2026agentforesight}
& -- & -- & \checkmark & -- & -- & -- \\

PrefixGuard~\cite{huang2026prefixguard}
& -- & -- & \checkmark & -- & -- & -- \\

HINTBench~\cite{wang2026hintbench}
& -- & -- & -- & -- & -- & -- \\

\textbf{Ours}
& \textbf{\checkmark}
& \textbf{\checkmark}
& \textbf{\checkmark}
& \textbf{\checkmark}
& \textbf{\checkmark}
& \textbf{\checkmark} \\
\bottomrule
\end{tabular}%
}
\end{table*}

We organize the single-agent literature with the three-stage taxonomy of BackdoorAgent~\cite{feng2026backdooragent}, which partitions an agent's attack surface into \emph{planning}, \emph{memory}, and \emph{tool use} and shows that high attack success is achievable at all three. Its proposed defense, comparing clean and triggered output-token distributions, is insufficient: because the malicious effect is spread across many interleaved benign reasoning steps, this dilution weakens the token-distribution signal that defense relies on. This dilution across steps is what motivates trajectory-level analysis, and it returns in a sharper form in the distributed attack we study.

\paragraph{Planning.} Planning attacks target the reasoning process, introducing a malicious instruction or step that redirects the agent's decisions without touching the environment or the model weights. The trigger may sit in the prompt, needing only black-box access, as in BadChain~\cite{xiang2024badchain}, or be implanted during fine-tuning so that an environmental cue activates attacker-chosen behavior later at inference time, as in BadAgent~\cite{wang2024badagent}. The compromise lives in the model's own reasoning trace rather than in external data.

\paragraph{Memory.} Memory attacks poison the retrieval layer that deployed agents depend on. A few adversarial documents that contradict the correct answer can suffice~\cite{zou2025poisonedrag}; other work encodes the trigger-response mapping in the retrieval index itself so the poisoned documents remain semantically coherent~\cite{cheng2024trojanrag}; and when the memory is an agent's own experience store, gradient-based optimization over the embedding space yields trigger phrases under very small poisoning budgets~\cite{chen2024agentpoison}. The attack surface is the data returned at retrieval time, not the model.

\paragraph{Tool use.} Tool-use attacks inject malicious content into the observation channel, the raw output tools return after each action, which the agent cannot distinguish from a legitimate response. AdvAgent~\cite{xu2024advagent} redirects web agents through adversarially crafted tool responses, even against closed commercial systems, and DemonAgent~\cite{zhu2025demonagent} embeds dynamically encrypted backdoor fragments in observations, using rotating trigger-response mappings that defeat static signatures, though it remains single-agent. The high-entropy ciphertext DemonAgent injects is the same kind of signal our detector uses, and this is the stage our own attack occupies.

None of these attacks crosses agent boundaries, all of them are based on systems that rely on a single agent performing the task from the beginning to the end. Collaborative Shadows~\cite{zhu2025collabshadows} is, to our knowledge, the first attack to spread an executable payload across the agents of a multi-agent LLM system: it distributes a payload across tool observations and agents, so that no single agent ever holds the complete payload, and an external step reassembles and executes it after the run completes. Each observation contains at most one uninterpretable fragment. A per-step check therefore cannot reconstruct the attack as a whole. This is the structural gap we study. Recent work shows that local monitoring can fail once attack fragments become indistinguishable from benign traffic in the monitored view~\cite{hu2026localmonitors}. Our fragments remain locally visible, so here we instead ask how early that signal can be used. 

The closest attack in mechanism is SkillTrojan~\cite{feng2026skilltrojan}, which likewise partitions an encrypted payload across several benign-looking invocations and reconstructs it under a trigger, but it operates through skill composition inside a single code-based agent rather than across cooperating agents, and it reports attack success and utility without any detection analysis. Backdoors in cooperative multi-agent reinforcement learning also distribute a trigger over time~\cite{yu2024spatiotemporal,zhu2026tresbd}, but in a control setting unlike tool-using LLM agents.

Because per-step audits miss patterns that appear only across a run, a growing line of work evaluates safety at the trajectory level. R-Judge~\cite{yuan2024rjudge} tests whether LLMs can identify risks from trajectory context; ATBench with its AgentDoG guardrail~\cite{liu2026agentdog} adds root-cause diagnosis under a source/failure-mode/consequence taxonomy; OpenAgentSafety~\cite{vijayvargiya2026openagentsafety} executes real tools under benign and adversarial intent; and HINTBench~\cite{wang2026hintbench} studies risk that arises with no attacker present, from the agent's own behavior drifting over a long task. A parallel line measures how one agent's answer shifts when other agents disagree~\cite{hu2026most,qu2026easier,hu2026one}. Prompt-injection and tool-risk benchmarks such as InjecAgent~\cite{zhan2024injecagent}, AgentDojo~\cite{debenedetti2024agentdojo}, and ToolEmu~\cite{ruan2024toolemu} evaluate tool-integrated agents but define attacks as test-case scenarios rather than logged runs. Across this landscape the unsafe event is a visible behavioral failure or an injected instruction, never a covert payload that assembles across agents, and none of these resources releases multi-agent trajectories under a distributed backdoor.

Our timing analysis is closest to recent work on online, prefix-level risk prediction. AgentForesight~\cite{zhang2026agentforesight} reframes failure attribution as online auditing, alarming at the earliest decisive error from the current prefix, and PrefixGuard~\cite{huang2026prefixguard} learns, from raw execution traces, a runtime monitor that scores the risk of the run so far. We share their premise, that a partial trajectory can carry a risk signal. 

The difference is what is being predicted. These systems forecast \emph{task failure}: the agent makes a mistake that a task verifier can score, and their labels come directly from that verifier. Our unsafe event is the opposite situation, an attacker succeeding while the agent completes its benign task normally. There is no task verifier that flags it: the payload is encrypted, the run looks successful, and the fragments are split across agents so no single observation contains the complete malicious payload. Our labels instead come from injection, assembly, and execution landmarks that we instrument into the attack. We therefore ask how early this hidden adversarial event becomes visible on a real attack. Table~\ref{tab:related_work_comparison} summarizes the comparison: ours is the only work that combines a distributed and actually executed payload with online prefix monitoring, and the only one to instrument attack landmarks, measure the pre-assembly window they define, and test robustness once those visible cues are reduced.

\section{Threat Model, Attack, and Corpus}
\label{sec:methodology}

\begin{figure*}[t]
\centering
\includegraphics[width=\textwidth, keepaspectratio]{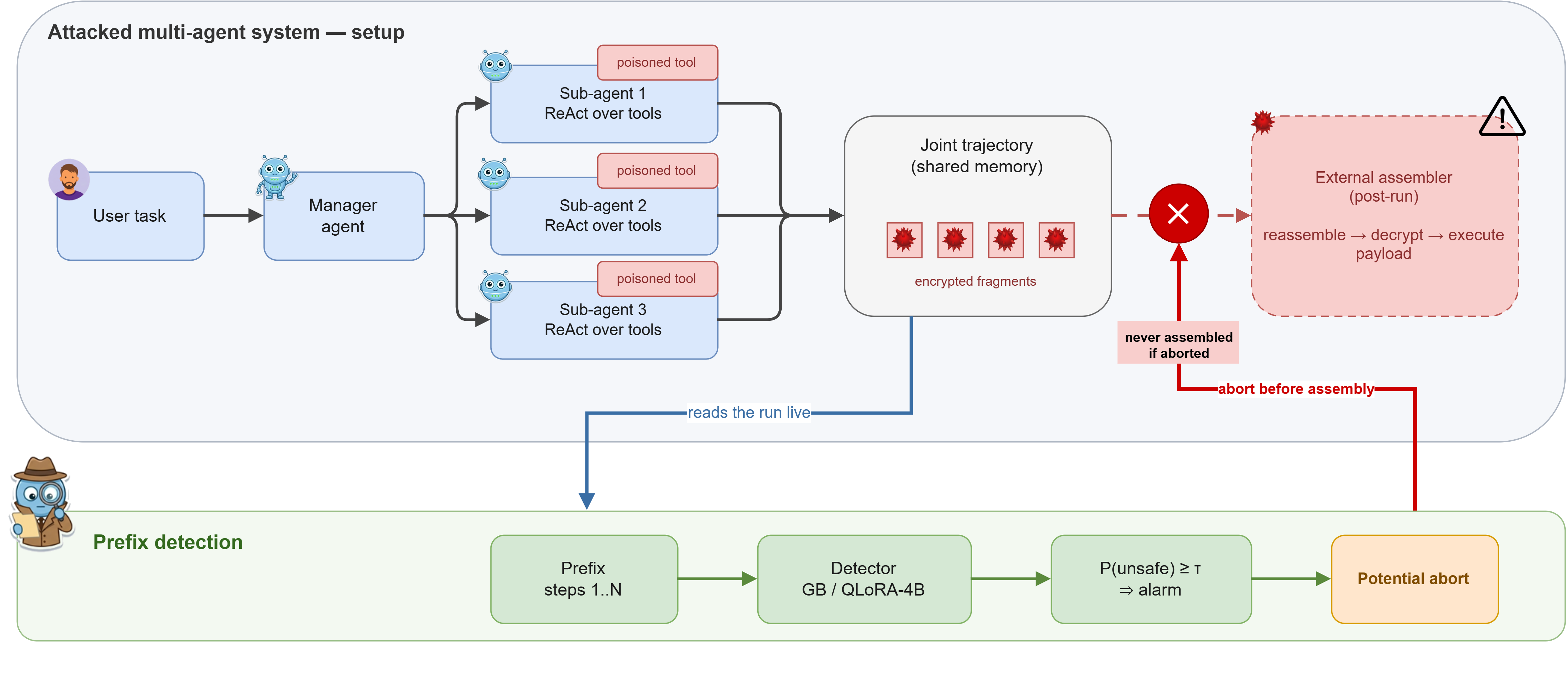}
\caption{Framework overview. A poisoned multi-agent system (top) hides encrypted payload fragments in tool observations that accumulate across the joint trajectory; only after the run does an external assembler reconstruct and execute them. Our prefix detector (bottom) reads the trajectory as it unfolds and, once its unsafe probability crosses~$\tau$, raises an alarm that could trigger an abort before the assembler executes the payload.}
\label{fig:framework}

\end{figure*}

\paragraph{Threat model.} The attacker controls a subset of the tool wrappers and a post-run assembly component, and issues the user task the system carries out; they cannot modify the agents, their prompts, the language models, the centralized monitor, or the enforcement mechanism. This lets them plant encrypted payload fragments in tool observations and share the decryption key. Their goal is to have an arbitrary payload assembled and executed from fragments that are individually harmless, with no single agent ever holding the complete payload and no isolated step revealing the complete attack. The defender sees one event stream covering every agent's thought, action and observation as the run unfolds. It does not see the external assembler. It can stop the run before the agents' memory reaches the assembler. Its aim is to flag an attacked run from that stream, and ideally to do so before the payload is assembled and executed. Our timing result therefore applies specifically to systems in which post-run processing is gated by a trusted enforcement layer.

\subsection{The Attack: Collaborative Shadows}
\label{sec:attack}

The attack targets a hierarchical multi-agent system. A \emph{manager} agent receives the user's task, decomposes it into sub-tasks, and dispatches each to one of several specialized \emph{sub-agents}. Each sub-agent runs its own ReAct loop~\cite{yao2023react} over a domain-specific set of tools, and once every sub-agent has finished the manager aggregates their results into a final answer. From the user's point of view the system behaves as an ordinary assistant (Figure~\ref{fig:framework}).

A subset of the tools are \emph{poisoned}. When a poisoned tool is called, its wrapper runs the genuine underlying tool, takes the real observation, and appends an encrypted fragment of a Python payload behind a delimiter before returning it. The payload is encrypted with AES-256-CBC. The key comes from the task text the attacker issued, so only the attacker can decrypt it. Each fragment is one slice of the ciphertext, so no fragment can be read on its own. Each poisoned tool carries one fragment. The observation, ciphertext included, is fed straight back into the agent's next reasoning step. 

The attack does not try to hide the fragment from the agent, it relies on the encrypted suffix being noise that carries no instruction the model would act on. This is the main pillar of this attack: the agent's chain of thought proceeds normally while its memory quietly accumulates fragments of malicious code.

After the system returns its answer, an assembly step that runs outside the agents sweeps their combined memory, recovers every fragment, decrypts them, and reconstructs the payload in order. If the reconstructed payload is complete and valid, it is executed. This gives each run a simple timeline. Fragments are injected one at a time while the run is still going, but assembly and execution each happen once, after the final step. A detector watching a live run can therefore see injection as it occurs, but not assembly or execution, which have not happened yet. How early a detector can react, and how that timing sits relative to assembly and execution, is what we measure.

Our version is a fresh re-implementation of Collaborative Shadows~\cite{zhu2025collabshadows} in our own codebase, and it differs in two ways that matter for detection. First, the sub-agents run concurrently rather than one after another. Second, we instantiate the attack under two structurally different tool environments, which gives two distinct fragment distributions and two very different observation styles. In \texttt{database\_searching}, three sub-agents (Employees, Projects, Salaries) query a relational SQLite database and observations are structured tabular text; in \texttt{api\_tools}, three sub-agents (Weather, News, Geo) call API-style tools and observations are JSON payloads with numeric and geographic fields. In each environment four fragments are spread across four tool calls, and all four must be recovered for the payload to assemble.

\subsection{Corpus}
\label{sec:corpus}

We generated the corpus by running the attack system against a fixed task set under two conditions, clean and attacked, for five language models and both tool environments. The models are \texttt{gemma4:26b}, \texttt{qwen3.5:35b}, \texttt{qwen3:4b}, \texttt{qwen2.5:7b-instruct}, and \texttt{llama3.2:3b}, all served locally through Ollama, and each model, environment, and condition combination was run over one hundred tasks, with some
 additional runs collected during data generation, giving 2{,}082 trajectories after discarding one that ended in an error. Runs average 14 steps. The two conditions differ only in whether the tool wrappers are poisoned and able to inject fragments; the agents receive identical instructions. An attacked run in which the agents happened to call no poisoned tool carries no fragment and counts with the uninjected runs below.

Every run is logged as a sequence of (thought, action, observation) steps for the manager and its sub-agents. The logs are written in neutral form: no field name or metadata field encodes the run's condition, so a trajectory can be handed to an external detector without leaking its label. Alongside each run we record the landmarks the study depends on, the step at which each fragment is injected and, for attacked runs, whether reassembly and execution succeed once the run has finished. A run is labeled \emph{unsafe} only when reassembly and execution both succeed, and \emph{safe} otherwise. We keep this ground-truth label distinct from whether fragments were merely present, since an injected run can still fail to assemble. 

The 2{,}082 runs split three ways. In 1{,}101 runs no fragment was ever injected. In the remaining 981 runs the poisoned tools injected payload fragments; of these, 553 failed to reassemble or execute after the run, while in 428 the payload was reassembled and executed. Since a run is labeled by outcome, the 1{,}101 uninjected runs and the 553 injected-but-failed runs are all safe, giving 1{,}654 safe runs (79.4\%) against 428 unsafe. 

Separately from these 2{,}082 runs, we also generated a held-out set of 300 additional trajectories from entirely new tasks, used only for evaluation. The corpus is dominated by safe runs, both because half the runs are clean and because many injected attacks never assemble, and attack success varies sharply across models. Table~\ref{tab:asr} breaks attack success rate down by model and tool environment. The two do not act independently: some models succeed far more often under \texttt{api\_tools} while others do so under \texttt{database\_searching}, so susceptibility is a property of the model and the tool environment together rather than of model scale alone.

\begin{table}[t]
\centering
\caption{Attack success rate (\%) by model and tool environment, over injected runs.}
\label{tab:asr}
\setlength{\tabcolsep}{4pt}
\resizebox{\columnwidth}{!}{%
\begin{tabular}{lccc}
\toprule
Model & \texttt{api\_tools} & \texttt{database\_searching} & Overall \\
\midrule
\texttt{gemma4:26b} & 62.7 & 80.0 & 72.1 \\
\texttt{qwen3.5:35b} & 54.0 & 70.0 & 62.3 \\
\texttt{qwen3:4b} & 84.0 & 19.0 & 51.5 \\
\texttt{llama3.2:3b} & 35.0 & 5.0 & 20.1 \\
\texttt{qwen2.5:7b-instruct} & 22.0 & 7.0 & 14.5 \\
\midrule
All models & 51.2 & 36.3 & 43.6 \\
\bottomrule
\end{tabular}%
}
\end{table}

Finally, the payload in the attack above is encrypted, so its fragments are high-entropy base64 text that stands out from ordinary tool output. This makes the attack easy to flag on surface statistics alone, and it leaves open whether a detector that fires on attacked runs has learned anything about the attack or is only reacting to the presence of ciphertext. To separate the two, we built a second variant of the corpus with a reversible \emph{word-codec} in place of the encrypted payload. A bijective map re-encodes the recorded Python fragments as ordinary English words. The injected text then looks like ordinary writing rather than base64, and it still decodes to the same payload. 

This variant is an offline retrofit of the logged trajectories, not a rerun of the agents. In each attacked run we replace the ciphertext suffix of every poisoned observation with the word-encoded form of the same fragment. The rest of the trajectory and the injection landmarks are left untouched. Every detector is then retrained and re-evaluated on the rewritten runs. We use this variant as a robustness check, to test whether detection survives once the entropy signature is removed and whether any surviving signal comes from the attack's structure rather than its surface.

\section{Experimental Setup}
\label{sec:setup}
\textbf{Detectors.} We evaluate six families of detector against the corpus. Two are used off the shelf: \emph{zero-shot LLMs} (\texttt{qwen2.5:7b-instruct} and \texttt{llama3.2:3b}) prompted to label a trajectory as safe or unsafe from its raw text, and \emph{AgentDoG}~\cite{liu2026agentdog}, a guardrail fine-tuned on the ATBench behavioral-safety corpus, applied to our runs without further training. Three are trained on our data: \emph{feature classifiers} (Logistic Regression, Random Forest~\cite{breiman2001randomforest}, and Gradient Boosting~\cite{friedman2001gradientboosting}) over content statistics of the observations; \emph{unsupervised anomaly detectors} (Isolation Forest~\cite{liu2008isolationforest}, One-Class SVM~\cite{scholkopf2001oneclasssvm}, and Local Outlier Factor~\cite{breunig2000lof}) fit on clean runs only, all implemented with scikit-learn~\cite{pedregosa2011scikit}; and a \emph{QLoRA-fine-tuned Qwen3-4B} that reads the raw thought, action, and observation text directly. A sixth is a deliberately minimal baseline, a \emph{per-observation detector} that scores each tool observation on its own, described below. 

\textbf{QLoRA training.} The fine-tuned model uses 4-bit NF4 quantization~\cite{dettmers2023qlora} with rank-16 low-rank adapters~\cite{hu2021lora} ($\alpha=32$). Training runs 3 epochs with a cosine schedule (learning rate $2\times10^{-5}$, 5\% warmup) at an effective batch size of 8 (per-device batch 1, gradient accumulation 8), with sequences capped at 3{,}072 tokens (truncating 4.4\% of examples) and minority-class oversampling to roughly 1:1. The pooled split is a single 80/20 train/validation partition; the cross-domain splits, between the two tool environments, train on one full environment and evaluate on the other. We use the adapter saved at the end of the final epoch. Training took $4.76$~hours in total on a single A100.

\textbf{Features and prefix protocol.} The feature classifiers read a fixed vector computed over the first $N$ steps of a run. We evaluate two feature sets: a 10-feature set that includes observation and thought length, and a 22-feature content set, with length features removed, built around observation Shannon entropy and base64-character ratio. To study detection over time we score each run (5-fold cross-validation) and read the Gradient Boosting probability off growing prefixes: at prefix length $N$ we compute the feature vector from only the first $N$ steps and read off the predicted unsafe probability. 

Training on complete runs mirrors deployment, where labels exist only once a run has finished, and the features are prefix aggregates computed the same way at every $N$. Scoring a prefix with a model trained on full runs is still a train-test shift. We also train a detector on all prefix lengths pooled, which matches training to scoring, and report its timing results with the original protocol shown alongside. The original detector fires later than the detector trained on prefixes. 

Sweeping $N$ upward, the first $N$ at which that probability exceeds a threshold $\tau$ is the run's first-detection step $N^\ast$; we report results at $\tau \in \{0.5, 0.75, 0.9\}$.

\textbf{Per-observation baseline.} To ask whether earlier steps are needed at all, we also evaluate a detector that sees only the current observation. At position $N$ it reads the $N$-th step and nothing before it. Its features are the observation's Shannon entropy, base64-character ratio and non-ASCII ratio, optionally with the observation's length. It is trained on the same run-level labels the prefix detector receives, and a run fires at the first step whose score crosses threshold, under the same alarm rule. The 22-feature content set excludes length by construction, so we run both detectors with and without an observation-length feature.

\textbf{Comparison at the same false-alarm rate.} Because the detectors produce different scores, a common probability threshold does not compare them fairly. Where we compare detectors directly we fix a false-alarm budget on the 1{,}101 uninjected runs and, for each detector, take the most sensitive threshold that stays within it. Each fold's threshold is set on the training folds only, and the realized false-alarm rate is reported on the held-out fold.

\textbf{Metrics.} We report accuracy, precision, recall, and $F_1$ for the label decisions, and AUROC (area under the ROC curve; $1$ is perfect separation, $0.5$ is chance) where a detector produces a ranking score. The reference point throughout is the majority-class baseline, which always predicts safe and reaches 79.4\% accuracy on the corpus's roughly four-to-one safe/unsafe split; a detector is useful only if it beats this.

\textbf{Evaluation splits.} We evaluate under three regimes of increasing difficulty. \emph{Pooled}: train and test are drawn from the same distribution (5-fold cross-validation over all 2{,}082 runs). \emph{Held-out tasks}: train on the 2{,}082 runs and test on 300 additional trajectories generated from 15 entirely new tasks per model, tool environment, and condition, with no task overlap. \emph{Cross-domain}: train on one tool environment and test on the other in both directions (\texttt{api\_tools}, 982 runs $\leftrightarrow$ \texttt{database\_searching}, 1{,}100 runs).

\textbf{Timing landmarks.} Three events anchor the timing analysis. \emph{Injection} is a per-step, in-run event: the step at which a poisoned tool first writes a fragment to an agent's memory. \emph{Assembly} and \emph{execution} are single events that occur once, after the run's final step. For the in-run timing analysis we also mark the step at which the fourth and final distinct fragment lands, the first moment the full payload is present in memory. No detector observes any landmark. We classify each detection as before, at, or after injection, and separately as before or after assembly. That tells us how early a run is flagged, and whether it is flagged in time to abort before the payload is reconstructed.

\section{Results}
\label{sec:results}

\subsection{The Attack Leaves an Actionable Warning Window}
\label{sec:earlywarning}
The central timing result is that the attack can be caught early, but only after it begins. We score each unsafe run on its growing prefix with the Gradient Boosting classifier (5-fold cross-validation), recording the first step at which the predicted unsafe probability crosses a threshold. Almost no run is flagged before its first fragment is injected. For the detector trained on full runs the pre-injection rate is $0.5\%$ of detected runs at the threshold we report; for the prefix-trained detector it is $0.9\%$, and for the per-observation detector it is $0\%$. Before injection there is no fragment to react to, so an earlier alarm would be a guess rather than a detection.

Once a fragment is present, detection follows quickly. At the most permissive threshold ($\tau=0.5$) the classifier flags $99.1\%$ of unsafe runs, a median of six steps after the first injection (median first detection at step~8). Training the same detector on prefixes instead of on complete runs moves first detection to the opening step, which leaves a median of twelve steps before the run ends. We call this the \emph{prefix-trained} detector and use it for the headline timing result; Table~\ref{tab:timing} gives its rates. Raising the threshold on the full-run-trained detector trades coverage for fewer false alarms: at $\tau=0.9$ it still catches $86.0\%$ of runs, but a median of eleven steps in (Table~\ref{tab:timing}). 

\begin{table}[t]
\centering
\caption{Detection timing and false-alarm cost on the ciphertext corpus, scored out of fold over the 428 successful attacks. }
\label{tab:timing}
\setlength{\tabcolsep}{3pt}
\resizebox{\columnwidth}{!}{%
\begin{tabular}{lcccccc}
\toprule
 & Detection & Median & Steps & \multicolumn{3}{c}{False alarms} \\
\cmidrule(lr){5-7}
$\tau$ & rate & $N^\ast$ & remaining & uninj. & inj.-failed & pooled \\
\midrule
0.5 & 99.1\% & 8 & 5 & 0.4\% & 30.7\% & 10.5\% \\
0.75 & 94.9\% & 9 & 4 & 0.2\% & 15.2\% & 5.2\% \\
0.9 & 86.0\% & 11 & 3 & 0.2\% & 7.1\% & 2.5\% \\
\midrule
\multicolumn{7}{l}{\emph{Same detector, retrained on prefixes, at a 1\% uninjected false-alarm budget}} \\
-- & 99.5\% & 1 & 12 & 1.1\% & 80.5\% & 27.6\% \\
\bottomrule
\end{tabular}%
}

\end{table}

What that lead is worth depends on which event a defender wants to prevent. The assembler runs only after the manager returns, so any alarm raised while the run is still executing comes before assembly. Concretely, first detection leaves a per-run median of twelve steps before the run ends, and because assembly runs only after that final step, those remaining steps are the window in which an automatic abort could stop the payload from being reconstructed. At a $1\%$ uninjected false-alarm budget the detector flags $99.5\%$ of the attacks. On almost every successful attack the alarm therefore arrives while there is still time to stop the run, however many fragments have already piled up in memory. The detector is post-injection but pre-assembly: it cannot warn that an attack is coming, but it can leave time to intervene before reconstruction.

An early warning is only useful if it is not drowned in false alarms, so the last three columns of Table~\ref{tab:timing} report what the detector costs on safe runs. Safe runs come in two kinds, and they behave differently. Almost all the alarms fall on the runs that carried fragments but failed to reassemble; runs in which no fragment was ever present are rarely flagged. An injected-but-failed run does carry a payload; it simply never assembled. We report the two rates separately, with the pooled rate beside them, for both protocols. The prefix-trained detector costs $1.1\%$ on uninjected runs and $27.6\%$ pooled. Raising the confidence threshold buys a quieter detector at the price of later, rarer detection. The warning is best read as a trade-off between lead time and false alarms, not a single threshold.

\begin{figure*}[t]
\centering
\includegraphics[width=0.92\textwidth, keepaspectratio]{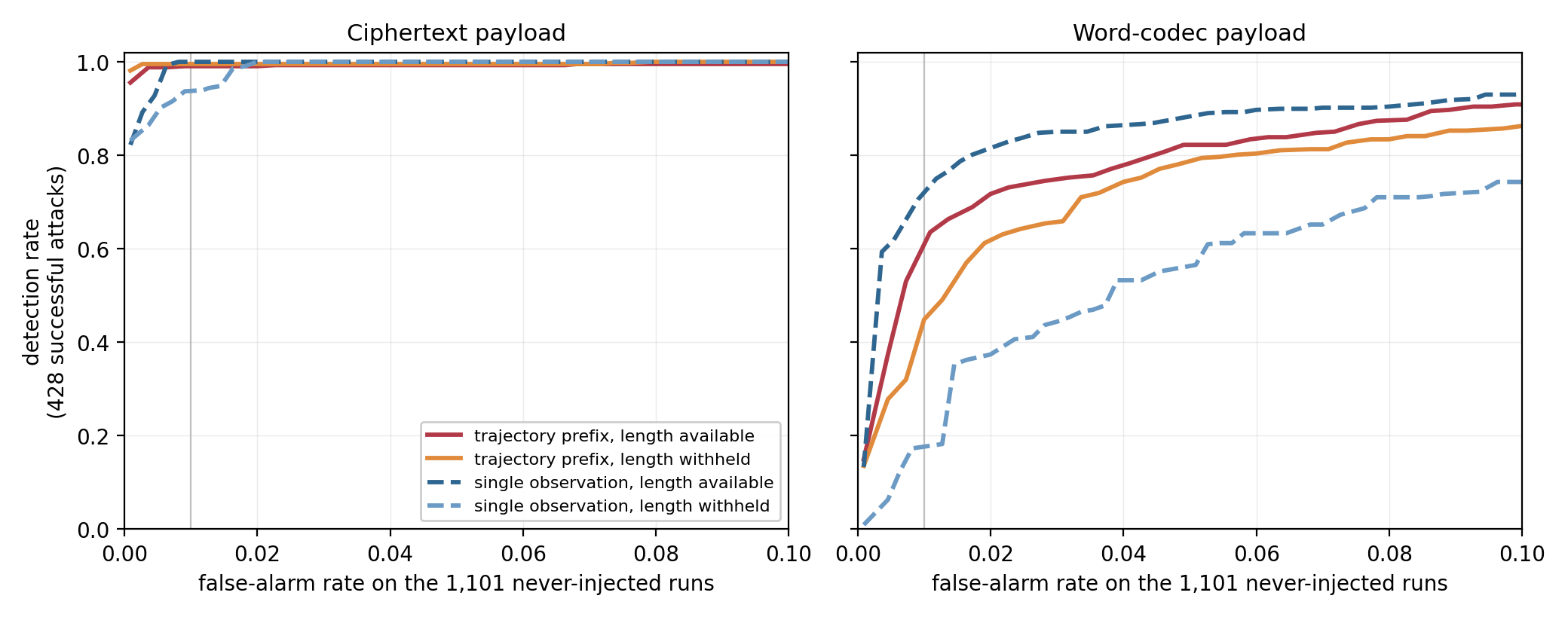}
\caption{Detection against false-alarm cost for both views, with and without an observation-length feature. On the ciphertext corpus (left) the curves are indistinguishable above a $2\%$ false-alarm rate. On the word-codec corpus (right) they separate, and the separation turns on whether length is available: the per-observation view is the higher of the two with it ($77.8\%$ against $65.2\%$) and the prefix view catches considerably more without it. Curves sweep the threshold directly, so they do not depend on a calibration choice; the vertical line marks the $1\%$ budget at which Table~\ref{tab:context} is read.}
\label{fig:context}
\end{figure*}
\subsection{How Much Context Is Needed?}
\label{sec:context}

Section~\ref{sec:earlywarning} scores the prefix detector at every window length. It leaves one question open: how well does a detector do when it sees only the current observation? Answering it tells us whether the warning needs the earlier steps at all. We compare the two views at a matched false-alarm budget, each with and without an observation-length feature.

On the ciphertext corpus the two views perform almost identically. At a $1\%$ budget every variant flags between $93.7\%$ and $100\%$ of the 428 successful attacks, and all of them fire within the first two steps (Table~\ref{tab:context}, Figure~\ref{fig:context}). An AES-encrypted, base64-encoded blob appended to ordinary tool output is conspicuous in the step that carries it, so a detector gains little by waiting for more steps.

The word-codec corpus separates them, and the separation turns on one feature. With observation length available to both, the per-observation detector is the better of the two, $77.8\%$ versus $65.2\%$. With length withheld from both, the ordering reverses: the prefix detector reaches $44.4\%$ versus the per-observation detector's $30.4\%$, pooled across the two tool environments. When length is removed, using earlier steps improves detection overall, though the gain is not uniform across the two environments.

\begin{table}[t]
\centering
\caption{How much context is needed, over the 428 successful attacks at a matched $1\%$ false-alarm budget.}
\label{tab:context}
\setlength{\tabcolsep}{4pt}
\resizebox{\columnwidth}{!}{%
\begin{tabular}{llcccc}
\toprule
 & & \multicolumn{2}{c}{Ciphertext} & \multicolumn{2}{c}{Word-codec} \\
\cmidrule(lr){3-4}\cmidrule(lr){5-6}
View & Length & Detection & $N^\ast$ & Detection & $N^\ast$ \\
\midrule
Trajectory prefix & available & 99.3\% & 1 & 65.2\% & 5 \\
Trajectory prefix & withheld & 99.5\% & 1 & 44.4\% & 6 \\
Per-observation & available & 100\% & 1 & 77.8\% & 5 \\
Per-observation & withheld & 93.7\% & 2 & 30.4\% & 5 \\
\bottomrule
\end{tabular}%
}

\end{table}

Breaking the word-codec comparison down by model and tool environment shows that the aggregate effect depends on the environment. With length withheld, the prefix detector is ahead on all five models in \texttt{database\_searching} and behind on four of five in \texttt{api\_tools}. Whether one observation is enough therefore depends on what the environment's ordinary output looks like.

Two properties of the protocol matter for reading these numbers. The per-observation detector never fires before the first fragment is injected, at any budget in either regime, so its early alarms track the attack rather than the task. And the realized uninjected rates are $1.0$--$1.7\%$, so the four conditions stay closely matched.

The detector fires early mostly because the injected text makes an observation longer, not because it puts evidence together across agents. Take the length away and the prefix detector does better than the per-observation detector. A detector that reads the run so far should therefore always be compared with one that reads only the current observation. None of this weakens the attack: no single observation ever holds the payload. One observation is enough to notice that a payload is being planted, which is a smaller claim.

\subsection{Generic Safety Detectors Fail, Payload Surface Cues Enable Detection}
\label{sec:signal}
The strongest signal available to our detectors is a surface property of this attack: its high-entropy ciphertext and the length that ciphertext adds, rather than the way the payload is distributed across agents. The baselines that watch for bad behavior do not separate unsafe from safe runs. Prompted to judge a trajectory from its raw text, two zero-shot LLMs fail in opposite directions: \texttt{qwen2.5:7b-instruct} calls almost everything safe ($F_1=0.019$) while \texttt{llama3.2:3b} calls most things unsafe ($F_1=0.280$). AgentDoG~\cite{liu2026agentdog}, a guardrail trained on behavioral-safety trajectories, fires on $91.8\%$ of runs indiscriminately for $25.9\%$ accuracy, below the majority-class baseline (Table~\ref{tab:baselines}). All three fail for the same reason: the agents' reasoning is behaviorally normal at every step, so a detector that looks for behavior anomalies has nothing to see.

\begin{table}[t]
\centering
\caption{Baseline detectors on the full ciphertext corpus, with 5-fold cross-validation for the trained rows.}
\label{tab:baselines}
\setlength{\tabcolsep}{4pt}
\resizebox{\columnwidth}{!}{%
\begin{tabular}{lrrrr}
\toprule
Detector & Acc. & Prec. & Rec. & $F_1$ \\
\midrule
Zero-shot \texttt{qwen2.5:7b-instruct} & 75.6\% & 0.056 & 0.012 & 0.019 \\
Zero-shot \texttt{llama3.2:3b} & 40.3\% & 0.186 & 0.563 & 0.280 \\
AgentDoG~\cite{liu2026agentdog} & 25.9\% & 0.209 & 0.932 & 0.341 \\
\midrule
RF, 10 features (with length) & 87.9\% & 0.740 & 0.642 & 0.686 \\
GB, 10 features (with length) & 88.5\% & 0.732 & 0.701 & 0.714 \\
RF, 8 features (length removed) & 78.1\% & 0.324 & 0.072 & 0.116 \\
\bottomrule
\end{tabular}%
}
\end{table}

Simple feature classifiers do beat the baseline. Random Forest and Gradient Boosting over ten cheap features reach $F_1=0.686$ and $0.714$, yet for Gradient Boosting a single feature, the average observation length, carries about $60\%$ of the importance, and thought length for most of the rest. The encrypted payload physically lengthens the observations that carry it, so the classifiers are really measuring payload size. Removing the two length features from the 10-feature set, leaving its eight remaining features (a set distinct from the 22-feature content set of Section~\ref{sec:setup}), confirms it: Random Forest collapses to $F_1=0.116$, and at $78.1\%$ accuracy it no longer beats the majority-class baseline.

With the length features removed, the question is what remains in the observation \emph{content}. The 22-feature content set of Section~\ref{sec:setup} answers it through two statistics: the Shannon entropy of the observation text and its base64-character ratio, the fraction of its characters drawn from the base64 alphabet. AES ciphertext encoded as base64 raises both well above the level of ordinary tool output, and a Gradient Boosting classifier over these content features reaches $F_1=0.886$ in-distribution with no access to any delimiter (Table~\ref{tab:generalization}). Both statistics are properties of a single observation, which is what Section~\ref{sec:context} exploits.

\begin{table}[t]
\centering
\caption{Gradient Boosting over the 22-feature content set, $F_1$ by evaluation split on the full ciphertext corpus.}
\label{tab:generalization}
\begin{tabular}{lr}
\toprule
Split & GB $F_1$ \\
\midrule
In-distribution (pooled) & 0.886 \\
Held-out tasks & 0.891 \\
Cross-domain api$\to$db & 0.000 \\
Cross-domain db$\to$api & 0.647 \\
\bottomrule
\end{tabular}

\end{table}

That signal, however, is tied to one domain's text. Trained on \texttt{api\_tools} and tested on \texttt{database\_searching}, the same classifier scores $F_1=0.000$. It learned where to put its threshold from \texttt{api\_tools} text. \texttt{database\_searching} output looks different enough that the threshold sits in the wrong place, though the classifier still ranks unsafe runs above safe ones. The reverse direction is easier but still degraded (db$\to$api $F_1=0.647$), while held-out tasks \emph{within} a domain transfer fine ($F_1=0.891$). The failure is specific to the cross-domain shift, not to unseen tasks.

If the ciphertext is an anomaly absent from clean runs, an outlier model fit on clean data alone should flag it. It does not. Isolation Forest, One-Class SVM, and Local Outlier Factor all sit at or below chance across domains. On api$\to$db every AUROC falls below the $0.5$ of random guessing, so these detectors rank unsafe runs as more typical than safe ones. None clears $0.7$ even pooled. Averaging content features over the whole trajectory dilutes a signature that lives in one or two steps, and without ever seeing an attack the models cannot calibrate which deviation matters. A different encoding, or a behavioral attack with no payload, may leave these detectors nothing to find.

\subsection{Reducing Surface Cues Delays Detection and Weakens Cross-Domain Transfer}
\label{sec:qlora}
\label{sec:robustness}
 Section~\ref{sec:signal} left the exploitable signal resting on surface cues read through fixed statistics. This section stress-tests that dependence from both sides. From the detector side, we ask whether a model that learns from the raw trajectory text, rather than from a fixed feature set, degrades more gracefully; from the attack side, we remove the cues themselves, re-encoding the payload so its entropy signature disappears, and measure what detection and timing survive. 
 
 We ask the detector's side first: does a model fine-tuned on our own trajectories recover the cross-domain case that the feature detector loses?

 We fine-tune Qwen3-4B with QLoRA (Section~\ref{sec:setup}) to label a run from its raw thought, action, and observation text, and compare it against the best feature detector, Gradient Boosting over the 22-feature content set (Table~\ref{tab:qlora}). The two in-distribution numbers use different resampling protocols, 5-fold cross-validation for Gradient Boosting and a single 80/20 split for QLoRA. It recovers much of the gap. In-distribution the two are close ($F_1=0.905$ versus $0.886$), but where Gradient Boosting collapses across domains (api$\to$db $F_1=0.000$), the fine-tuned model reaches $0.531$.

\begin{table}[t]
\centering
\caption{QLoRA-fine-tuned Qwen3-4B versus the best feature detector, ciphertext corpus.}
\label{tab:qlora}
\setlength{\tabcolsep}{4pt}
\resizebox{\columnwidth}{!}{%
\begin{tabular}{lrrrr}
\toprule
Split & GB $F_1$ & GB AUROC & QLoRA $F_1$ & QLoRA AUROC \\
\midrule
In-distribution (pooled) & 0.886 & 0.988 & 0.905 & 0.990 \\
Held-out tasks & 0.891 & 0.983 & 0.849 & 0.988 \\
Cross-domain api$\to$db & 0.000 & 0.864 & 0.531 & 0.886 \\
Cross-domain db$\to$api & 0.647 & 0.896 & 0.639 & 0.951 \\
\bottomrule
\end{tabular}%
}

\end{table}

The AUROC values sharpen what actually fails across domains. Both detectors still separate unsafe from safe runs on api$\to$db (GB $0.864$, QLoRA $0.886$), so neither loses the ranking signal; what breaks is the fixed decision threshold, which no longer sits in the right place once the domain changes. The difference between them is how well the threshold survives the shift: the fine-tuned model's threshold still yields a usable $F_1=0.531$, while Gradient Boosting's collapses to $0.000$. Working from the raw text rather than a fixed set of entropy statistics is consistent with the decision boundary degrading more gracefully in a domain never seen in training. 

Nonetheless this method is costly: training took $4.76$~hours where Gradient Boosting needs seconds, the db$\to$api direction barely moves ($0.639$ against $0.647$), and the model still relies on the ciphertext actually appearing in the text.

Both detectors, however, read ciphertext with two loud cues: high entropy and added length. We introduced both when we built the payload; neither is intrinsic to the attack. We therefore turn to the attack side and repeat the analysis on the word-codec corpus (Section~\ref{sec:corpus}), which re-encodes the recorded payload as ordinary words. The base64 cue is gone; the trajectory and the payload are unchanged.

With the entropy tell gone, we first asked what a feature detector keys on instead, expecting length to take over. It did: the word-encoded fragments still run about a third longer than clean output, and for a Gradient Boosting classifier observation length (\texttt{obs\_len\_max}) is the dominant feature by a wide margin ($0.58$ of the importance, against $0.07$ for the next). In-distribution it reaches $F_1 = 0.920$ (Table~\ref{tab:wordcodec-gb}).

We then removed the length features to see what remained. In-distribution the detector barely moved, to $F_1 = 0.861$, so length was not its only tell. What carries that residual, though, is another layer of \emph{surface} statistics rather than anything structural: with the length features removed, importance spreads across residual observation entropy and a set of character-level ratios (uppercase, non-ASCII, type-token), the fingerprints of wordy prose. 

They do not hold up in the harder api$\to$db direction. Training on api and testing on db, the detector scores $F_1 = 0.829$ with the length features but only $0.026$ without them: the length envelope was the one part of the statistical signal that held across environments. One structural feature survives the ablation. The fragments reuse tokens that are rare in ordinary tool output, and once length is gone that overlap is the Random Forest's top feature. It comes from the way the attack spreads its fragments over several observations, but on its own it is too weak to recover the cross-domain case.

\begin{table}[t]
\centering
\caption{Gradient Boosting on the word-codec corpus, with and without the explicit length features; $F_1$ on a balanced 600-run subset.}
\label{tab:wordcodec-gb}
\setlength{\tabcolsep}{4pt}
\begin{tabular}{lrr}
\toprule
Split & GB (full) & GB (no length feat.) \\
\midrule
In-distribution & 0.920 & 0.861 \\
api$\to$db & 0.829 & 0.026 \\
db$\to$api & 0.690 & 0.663 \\
\bottomrule
\end{tabular}

\end{table}

\begin{table}[t]
\centering
\caption{QLoRA-fine-tuned Qwen3-4B on the full word-codec corpus, without balanced subsampling.}
\label{tab:wordcodec-qlora}
\setlength{\tabcolsep}{4pt}
\begin{tabular}{lrr}
\toprule
Split & $F_1$ & AUROC \\
\midrule
In-distribution & 0.895 & 0.988 \\
api$\to$db & 0.457 & 0.854 \\
db$\to$api & 0.247 & 0.936 \\
\bottomrule
\end{tabular}

\end{table}

 The fine-tuned LLM recovers the harder cross-domain direction. Trained and tested on the word-codec corpus, it remains accurate in-distribution ($F_1 = 0.895$) and retains a usable threshold on api$\to$db ($F_1 = 0.457$, AUROC $0.854$; Table~\ref{tab:wordcodec-qlora}), the direction where the feature detector's threshold fails outright. The improvement is direction-dependent: on db$\to$api its $F_1$ falls to $0.247$ even as AUROC stays high ($0.936$), the threshold again sits in the wrong place. Because Table~\ref{tab:wordcodec-gb} is computed on the balanced subset and this table on the full corpus, we do not compare $F_1$ across the two tables directly. Scored on prefixes of the word-codec runs, it also flags the attack early, at a median first-detection step of $N^\ast = 2$ with a median delay of zero from the injection step (Table~\ref{tab:wc-timing}).
\begin{table}[t]
\centering
\caption{Prefix detection timing at $\tau = 0.5$ over the 428 successful attacks; median delay is measured from the injection step.}
\label{tab:wc-timing}
\setlength{\tabcolsep}{4pt}
\resizebox{\columnwidth}{!}{%
\begin{tabular}{llccc}
\toprule
Detector & Corpus & Detection rate & Median $N^\ast$ & Median delay \\
\midrule
GB & ciphertext & 99.1\% & 8 & 6 \\
GB, with length feat. & word-codec & 100\% & 7 & 6 \\
GB, no length feat. & word-codec & 98.6\% & 12 & 10 \\
QLoRA & word-codec & 100\% & 2 & 0 \\
\bottomrule
\end{tabular}%
}

\end{table}

The main test is whether the feature detector's own early-warning lead survives once the entropy envelope is gone, and whether it survives the further removal of length. We re-run that timing procedure unchanged on the word-codec corpus, with and without the length features (Table~\ref{tab:wc-timing}). Every row there uses the full-run-trained detector on the full corpus at $\tau=0.5$, so the two corpora are read under one protocol and the rates run higher than the budgeted figures of Table~\ref{tab:context}. With the length features present the lead survives the change of encoding almost exactly: median delay $+6$ after injection, reproducing the ciphertext result and so showing the early warning is not merely an artifact of the entropy envelope. 

After removing the length features from the detector, median first detection slips from step~7 to step~12 and the delay after injection grows from $+6$ to $+10$. The detector still flags nearly every successful attack before the run ends, and so still before the external assembler runs, but the alarm now typically arrives only once the payload is already complete in memory: only $22.3\%$ of runs are flagged before their fourth and final fragment lands. On ciphertext, length and entropy stand in for each other. The lead survives losing either one, but not both.

Taken together: stripping the entropy cue leaves detection intact, removing the length features from the detector delays it from step~7 to step~12 and collapses cross-domain transfer, and the fine-tuned model recovers part of that loss in the harder api$\to$db direction, though not in the reverse one.

\section{Discussion}
\label{sec:discussion}

The first sign of the payload appears at injection, so detection here can only react, never predict. That is enough. Assembly and execution happen only after the agents finish, so an alarm raised during the run still arrives in time, provided the enforcement layer gates what happens afterwards.

What decides the size of that window is the surface of the attack, not its distributed structure. Re-encoding the payload as ordinary words removes the entropy cue with little change in timing, so the warning is not an entropy artifact. Removing the length features as well pushes first detection several steps later and collapses cross-domain transfer: earlier steps help more once the injected text stops standing out.

Full-run accuracy is the wrong summary for an early-warning detector: it says nothing about whether the run could have been stopped in time. The prefix-trained detector's false-alarm rate is $1.1\%$ on uninjected runs, reported separately from runs that carried fragments but failed to assemble.

These results hold within clear limits. The warning window exists because assembly happens only after the run, at a point the enforcement layer can gate; an attack that reassembled and executed itself invisibly, with no such step, would remove the margin we rely on. Our controls reduce the surface cues but do not eliminate them: the trajectories are never length-matched, only the explicit length features are excluded from the detector, and even the word-codec setting still carries some residual lexical trace of the attack. And we study a single attack family across two tool environments, so what we can claim is about the reliability of this warning under domain shift, not about detection across other kinds of attack. Our fragments stay easy to spot in a single observation, because injection changes that observation's length and character statistics. An attack that hid those changes better would leave a per-observation detector with nothing to read. Finally, the word-codec corpus is an offline retrofit of the logged trajectories rather than a re-run of the agents under a word-encoded payload, so it isolates the cue without re-testing how agents behave when the injected text reads like ordinary writing. Making the attack fully covert and self-reassembling, and testing whether a detector built for the coordinated structure rather than the surface generalises further, are the natural next steps.

\section{Conclusion}
\label{sec:conclusion}

We set out to ask how early a distributed backdoor can be caught while a multi-agent run is still unfolding, and to answer it we built a working version of the attack, collected a corpus of full trajectories, and put a range of detectors against it. The attack can be caught, and caught in time. Almost no successful attack is flagged before its first fragment is injected. Once one is, a simple classifier reading only the run so far reacts early enough to abort the run before the payload is ever assembled, on almost every attacked run. The false-alarm cost is one the defender can tune against detection. 

How large that warning is, though, depends on what the injected text looks like. The early lead is carried in part by the ciphertext's length and entropy. Either cue alone is enough. Take both away and detection arrives later and transfers less reliably across tool environments, though a fine-tuned model recovers some of the loss. When the injected text still stands out on its own, reading one observation works as well as reading the whole run. The prefix detector pulls ahead only once we hide what makes the text stand out. The payload stays unrecoverable from any single step, which is what the attack was built for, but noticing that one is being planted does not need the whole run. The practical lesson is that an early-warning detector should be judged on lead time and false alarms rather than on full-run accuracy, and should be compared with a detector that sees only the current observation. Whether a truly covert, self-reassembling version of the attack can be caught, and whether a detector built for the coordinated structure rather than the surface would hold up under greater domain shift, are the questions we leave open.

\section*{Ethical Considerations}
This work builds and studies a working offensive capability, a distributed backdoor that hides an encrypted payload across the tools of a multi-agent LLM system and reassembles it after the run. We took the following positions to keep the work responsible. The base distributed attack follows the previously published Collaborative Shadows~\cite{zhu2025collabshadows}. We additionally study an offline word-encoded version of the recorded payloads to test detector robustness; we do not deploy this variant against live systems. 

All experiments ran in a closed, local sandbox: models were served locally, the tasks and databases are synthetic and contain no real personal data, and the reassembled ``payload'' is a controlled in-memory primitive that performs no real exfiltration and contacts no external system. No real API keys or endpoints appear in the corpus or in the released artifacts. No human subjects were involved.

The purpose and emphasis of the paper are defensive: we measure how early and how reliably such an attack can be flagged, so that operators of multi-agent systems can reason about detection and abort windows. The detection findings and the labeled trajectory corpus offer concrete value to defenders. 

\section*{LLM Usage Statement}Language models are also the object of study here. Five models act as the agents whose trajectories form our corpus, two are evaluated as zero-shot detectors, and one is fine-tuned as a detector. Sections~\ref{sec:corpus} and~\ref{sec:setup} give their identifiers, serving stack, training settings and evaluation protocol. Section~\ref{sec:discussion} says what these results do and do not show. Because the models are open-weight and served locally, the released corpus allows the detector experiments to be repeated without rerunning the agents.

The authors used generative AI tools to assist with drafting and editing prose and organizing results. They also used these tools to generate supporting analysis and visualization code. All experimental design, results, claims, and generated code were verified by the authors, who take full responsibility for the content of this paper. LLMs were used for editorial purposes in this manuscript, and all outputs were inspected by the authors to ensure accuracy and originality.

\section*{Acknowledgments}
This work used Jetstream2 at Indiana University through ACCESS allocation CIS260254 from the Advanced Cyberinfrastructure Coordination Ecosystem: Services \& Support (ACCESS) program, which is supported by U.S. National Science Foundation grants \#2138259, \#2138286, \#2138307, \#2137603, and \#2138296. We thank the Jetstream2 and ACCESS support teams for the computational infrastructure used in this work.

\bibliographystyle{IEEEtran}
\bibliography{bibliography}

\end{document}